\documentstyle[aps,preprint,tighten,floats,epsfig]{revtex}

\renewcommand{\vec}[1]{{\bf #1}}
\begin{document}
\preprint{}

\title{Examination of self interaction correction methods for Na clusters}
\author{K. Yoshizaki, A. Ono and N. Takigawa}
\address{Department of Physics, Tohoku University, Sendai 980-8578, Japan}
\maketitle

\begin{abstract}
  We examine whether the self interaction correction method by Harrison, 
  which does not introduce the spherical single particle density
  approximation to energy functional, can be applied to Na clusters.
  We show that it does not work well, especially, for large clusters,
  though it works well for atomic systems. 
  We suggest that it is better to apply this method only 
  to the Hartree term. 
  We also show that the effects of non-diagonal Lagrange multiplier
  originating from the orthonormality of single particle orbitals 
  are negligible. 
\end{abstract}


\section{Introduction}
\label{sec:intro}
The local density approximation (LDA) provides a 
powerful practical technique to apply the 
Kohn-Sham framework\cite{Koh65} to interacting many body problems.
A problem of this method is unphysical self interaction, 
\textit{i.e.} the interaction of a particle with itself.
Perdew and Zunger\cite{Per79,Per81}
proposed a prescription to remedy this shortcoming, 
which has been used for
atoms\cite{Per79,Per81,Har83a,Har83b},
molecules\cite{Ped84,Hea87},
bulk systems\cite{Hea83c,Har87} and
also metal clusters\cite{Mad95,Onw90,Pac93}.
Though there still remain some effects of self interaction, 
the major part of the problem is removed in this method.
Compared with the Hartree-Fock theory, 
which is free from the self interaction problem,
the local density approximation with the self interaction correction
(SIC) has advantages such as,
i)  exchange and correlation energies can be relatively 
easily handled in the same manner, 
ii) the resultant single particle energies well approximate
the physical removal energy from each orbit, 
iii) the numerical calculation is much lighter, especially 
for three dimensional calculations.

A characteristic feature of the SIC method of Perdew and Zunger 
is that the energy functional depends not only on the total 
density, but explicitly also on the density of 
each single particle orbital. 
In almost all calculations 
for closed shell atoms and metal clusters, 
the single particle densities in the energy functional are substituted 
by the spherically averaged densities. 
The central single particle potential is 
then deduced by taking functional derivative 
of the resultant energy functional 
with respect to the spherically averaged single particle density.
Following more closely the original idea of 
Perdew and Zunger, on the other hand, 
Harrison \cite{Har83a,Har83b} proposed a method 
of using the original single particle densities 
without introducing spherical averaging. 
Since the energy functional is not invariant 
under unitary transformation of single particle orbitals, 
Harrison represented the single particle orbitals by   
either spherical harmonics or Cartesian basis as two choices. 

Though the method by Harrison works well for atoms\cite{Har83a,Har83b},
it has not been tested for 
metal clusters. We address this question in this paper by 
taking Na clusters as an example.  
We show that it does not work well, especially for large clusters  
which have single particle orbitals with large angular momentum.
We confine our study to the exchange energy without referring 
to the correlation energy in order to make the argument clear and 
compare the results with those of Hartree-Fock calculations.
We show that a better agreement with the Hartree-Fock calculations 
is obtained if one applies Harrison's method only 
to the Hartree term. 

In addition to the validity of Harrison's method, we 
discuss in this paper the problem of 
non-diagonal Lagrange multipliers originating from the 
orthonormality of single particle 
orbitals in the self interaction correction method 
of Perdew and Zunger. 
We show that the non-diagonal property of the 
Lagrange multipliers introduces a negligible effect 
to single particle energies as well as the total energy. 

The paper is organized as follows. In Sec.\ref{sec:formalism} the 
SIC method of Perdew and Zunger and
Harrison's approach are briefly explained.
In Sec.\ref{sec:ene} the exchange energies calculated by several SIC
methods are compared, and the effect of non-diagonal Lagrange multipliers 
is discussed in Sec.\ref{sec:ortho}. 
Summary and conclusion are given in Sec.\ref{sec:summ}. 

\section{SIC formalism and Harrison's method}
\label{sec:formalism}
The total energy in the self interaction corrected
local density approximation (SIC-LDA) by Perdew and Zunger \cite{Per79,Per81} 
is expressed as
\begin{eqnarray}
  &&E_{\mathrm{TOT}}=T+E_{\mathrm{ext}}+E_{H}
  +E_{X}^{\mathrm{SIC}} \ , \label{eq:ETOT}
\end{eqnarray}
where
\begin{eqnarray}  
  &&T=-\frac{1}{2}\sum_{i}\int d^{3}r \ 
  \psi^{*}_{i}(\vec r)\nabla^{2} \psi_{i}(\vec r) \ ,\label{eq:kine}\\
  &&E_{\mathrm{ext}}[\rho]=\int d^{3}r \ 
  v_{\mathrm{ext}}(\vec r)\rho(\vec r) \ ,\\
  &&E_{H}[\rho]=\frac{1}{2}\int d^{3}r\int d^3{r'}
  \frac{\rho(\vec r) \rho(\vec r')}{\left|\vec r-\vec r' \right|} \ , \\
  &&E_{X}^{\mathrm{SIC}}=E_{X}^{\mathrm{LDA}}
  [\rho_{\uparrow}, \rho_{\downarrow}] 
  -\sum_{i=1}^{N}\left\{E_{H}[\rho_{i}]+E_{X}^{\mathrm{LDA}}[\rho_{i},0]
  \right\} \ , \label{eq:SICX}\\
  &&E_{X}^{\mathrm{LDA}}[\rho_{\uparrow}, \rho_{\downarrow}]=
  -\frac{3}{2}\left(\frac{3}{4\pi}\right)^{1/3}
  \sum_{\sigma =\uparrow, \downarrow}\int d^{3}r \ 
  \rho_{\sigma}(\vec r)^{4/3} \ .
\end{eqnarray}
Here all quantities are in Hartree atomic units, \textit{i.e.}
$m=e^{2}=\hbar=1$.
As mentioned in the introduction, the correlation energy has been neglected.
We take the effects of ions into account in the 
spherical jellium model. The external potential is then given by 
\begin{eqnarray}
  v_{\mathrm{ext}}(\vec r)&=&
  \left \{
    \begin{array}{ll}
      -Z/(2R_{\mathrm{jell}})
      \left\{3-\left(r/R_{\mathrm{jell}}\right)^{2}\right\}
      & r\le R_{\mathrm{jell}} \\
      -Z/r & r>R_{\mathrm{jell}} \ ,
    \end{array}
  \right.
\end{eqnarray}
where the jellium radius $R_{\mathrm{jell}}$ is related to the 
number of atoms in the cluster $Z$ by
$R_{\mathrm{jell}}=r_{s}Z^{1/3}$, $r_{s}$ being 
the bulk Wigner-Seitz radius which is 4 a.u. for Na.

The Euler equation under the orthonormality condition
\begin{eqnarray}
  \label{eq:Euler}
  \frac{\delta}{\delta \psi_{i}^{*}(\vec r)}
  \left \{E_{\mathrm{tot}}+\sum_{ij}\epsilon_{ij}\left(\delta_{ij}
      -\int d^{3}r \ \psi_{j}^{*}(\vec r)\psi_{i}(\vec r)\right)
  \right \}=0
\end{eqnarray}
results in the following coupled equations for the single particle wave 
functions
\begin{eqnarray}
  &&\left\{-\frac{1}{2}\nabla^{2}+v_{\mathrm{ext}}(\vec r)+\int d^3{r'}
    \frac{\rho(\vec r')}{\left|\vec r-\vec r' \right|}
    +v_{X}^{{\mathrm SIC}(i)}(\vec r)\right\}\psi_{i}(\vec r)
  =\sum_{j}\epsilon_{ij}\psi_{j}(\vec r) \ ,\label{eq:1eq}\\
  &&v_{X}^{{\mathrm SIC}(i)}(\vec r)=-\left(\frac{3}{\pi}\right)^{1/3}
  \rho^{1/3}(\vec r)
  -\left \{
    \int d^3{r'}\frac{\rho_{i}(\vec r')}{\left|\vec r-\vec r' \right|}
    -2\left(\frac{3}{4\pi} \right)^{1/3}\rho_{i}^{1/3}(\vec r)
  \right\} \ .
  \label{eq:vsicx}
\end{eqnarray}
The Lagrange multiplier $\epsilon_{ij}$ becomes non-diagonal because 
of the orbital dependence of the 
self interaction corrected exchange potential.
In the following, we approximate it by the diagonal components and 
solve the following non-coupled equations\cite{Per81}
\begin{eqnarray}
  &&\left\{-\frac{1}{2}\nabla^{2}+v_{\mathrm{ext}}(\vec r)+\int d^3{r'}
    \frac{\rho(\vec r')}{\left|\vec r-\vec r' \right|}
    +v_{X}^{{\mathrm SIC}(i)}(\vec r)\right\}\psi_{i}(\vec r)
  =\epsilon_{i}\psi_{i}(\vec r) \label{eq:diag1eq}
\end{eqnarray}
We discuss the validity of this approximation 
in Sec. \ref{sec:ortho} for Na clusters. 

As we see in Eq.(\ref{eq:SICX}) the SIC method of Perdew and Zunger 
is characteristic in that the total energy functional depends 
explicitly on the individual orbital density.
A consequence is that it loses 
invariance under the unitary transformation
of single particle orbitals. 

Another problem is that the numerical load is heavy, because 
one has to solve three-dimensional equations instead of
the one-dimensional 
equations for the radial motion of electrons even 
for closed shell atoms and metal clusters. 
In applying this formalism to those systems, 
one usually replaces 
$\rho_{i}(\vec r)$ in the curly brackets in Eqs.(\ref{eq:SICX}) and 
(\ref{eq:vsicx}) by the 
spherically averaged orbital density given by 
$\tilde{\rho}_{i}(r)$ 
\begin{eqnarray}
  \tilde{\rho}_{i}(r) &=&\frac{1}{4\pi}\int d \hat{\vec r} \ 
  \rho_{i}(\vec r) \ .\label{eq:rhoSA}
\end{eqnarray}
This prescription certainly reduces the numerical load, 
because the resultant single particle potentials become central 
potentials. However, it 
does not optimize the SIC following the original scheme 
of Perdew and Zunger. 

Harrison proposed an alternative procedure, which avoids 
replacing the single particle densities in the energy functional 
by spherically averaged ones. As mentioned  in the introduction, 
he expressed them in the spherical harmonic basis 
\begin{eqnarray}
  \rho_{nlm}^{\mathrm{SH}}(\vec r)=\left|\frac{u_{nl}(r)}{r}
    Y_{l}^{m}(\hat{\vec r})\right|^{2} \ , \label{eq:rhoSH}
\end{eqnarray}
or in Cartesian basis 
\begin{eqnarray}
  \rho_{nlm}^{\mathrm{C}}(\vec r)=\left\{
    \begin{array}{ll}
      \displaystyle
      \left|\frac{u_{nl}(r)}{r}
        Y_{l}^{0}(\hat{\vec r})\right|^{2} & m=0 \\
      \displaystyle
      \left|\frac{u_{nl}(r)}{r}\frac{Y_{l}^{|m|}(\hat{\vec r})
          \pm Y_{l}^{-|m|}(\hat{\vec r})}{\sqrt{2}}
      \right|^{2} & m\ne 0 \ .\label{eq:rhoC}
    \end{array}
  \right.
\end{eqnarray}
He noticed that the resultant energy functional can be expressed
in terms of the spherically averaged orbital density after the 
integration over angle,
\begin{eqnarray}
  E_{X}^{\mathrm{SIC}}&=&E_{X}^{\mathrm{LDA}}[\rho_{\uparrow}, \rho_{\downarrow}]
  -\sum_{n, l}2(2l+1)\left\{\sum_{k=0}^{2 l}
    E_{H}^{k,l}[\tilde{\rho}_{nl}]
    +c_{X}^{l}E_{X}^{\mathrm{LDA}}[\tilde{\rho}_{nl},0]\right\} \ ,
  \label{eq:mEX} \\
  E_{H}^{k,l}[\tilde{\rho}_{nl}]
  &=&\frac{1}{2}c_{H}^{k, l}\int dr \int dr'
  \left|u_{nl}(r)\right|^{2}
  \left|u_{nl}(r')\right|^{2}\frac{r_{<}^{k}}{r_{>}^{k+1}}  
  \ .\label{eq:mEH}
\end{eqnarray}
The coefficients $c_{H}^{k,l}$ and $c_{X}^{l}$ for
each of the spherical harmonic and Cartesian representations are listed
in Tables \ref{tab:coeffH} and \ref{tab:coeffX}.
Harrison then calculated the corresponding central potential 
for each set of quantum nembers $n,l$ 
by taking the functional derivative with respect to the spherically 
averaged orbital density $\tilde{\rho}_{nl}(r)$ 
\begin{eqnarray}
  v_{X}^{{\mathrm SIC}(nl)}(r)
  &=&\frac {\delta E_{X}^{\mathrm{SIC}}
    [\tilde{\rho}_{nl}]}
  {\delta \tilde{\rho}_{nl}(r)} \ .\label{eq:mv}
\end{eqnarray}

Though this method restricts the variational space smaller than 
that in the original scheme of Perdew and Zunger since it presumes a
spherically symmetric potential from the beginning, 
Harrison 
showed that his method still improves both the exchange energy and the 
total energy for atoms compared with the simple procedure 
where $\rho_{i}(\vec r)$ is replaced by the 
spherically averaged orbital density.  

It is an interesting question to see 
whether Harrison's method can be 
applied to metal clusters. 
A simple minded consideration would suggest that Harrison's method 
becomes more powerful in metal clusters.
This is because 
Harrison's treatment should have a large effect on high angular momentum 
orbitals which play more important roles in metal clusters than in atoms 
where the main contribution to the SIC originates from the $1s$ state
\cite{Per81}.
In the next section, we apply Harrison's method to 
Na clusters, and show that it does not work well contrary to 
the simple expectation. 
\begin{table}
  \begin{center}
    \begin{tabular}[H]{llrr} 
      $l$ & $k$ & spherical & Cartesian\\
      &     & harmonics & \\  \hline
      1 & 2 & 0.0800 & 0.1600  \\ 
      2 & 2 & 0.0571 & 0.0816  \\ 
      2 & 4 & 0.0317 & 0.0816  \\ 
      3 & 2 & 0.0533 & 0.0686  \\ 
      3 & 4 & 0.0202 & 0.0346  \\ 
      3 & 6 & 0.0179 & 0.0538  \\ 
      4 & 2 & 0.0519 & 0.0632  \\ 
      4 & 4 & 0.0180 & 0.0269  \\ 
      4 & 6 & 0.0108 & 0.0208  \\ 
      4 & 8 & 0.0119 & 0.0398  \\ 
    \end{tabular}
    \caption{The coefficients $c_{H}^{k, l}$ for the Hartree term.}
    \label{tab:coeffH}
  \end{center}
\end{table}

\begin{table}
  \begin{center}
    \begin{tabular}[H]{lrr}
      $l$ & spherical & Cartesian \\
      & harmonics & \\  \hline
      1 & 1.0937 & 1.1800 \\ 
      2 & 1.1293 & 1.2384 \\ 
      3 & 1.1508 & 1.2708 \\ 
      4 & 1.1659 & 1.2925 \\ 
    \end{tabular}
    \caption{The coefficients $c_{X}^{l}$ for the exchange term.}
    \label{tab:coeffX}
  \end{center}
\end{table}

\section{Results and discussion}
\label{sec:result}

\subsection{Exchange and total energy}
\label{sec:ene}
We compare in Table \ref{tab:EX} 
the exchange energy per electron for Na clusters 
calculated by several methods.
In this table and in what follows,
the abbreviations SA-SICX, SH-SICX and C-SICX stand for 
the SICX calculation using the spherically averaged, spherical harmonic and
Cartesian orbital densities, respectively.
The difference between the HF and the other calculations 
represents the error of each method since the HF calculation provides 
the exact exchange energy.
Strictly speaking, the exact exchange energy in the Kohn-Sham formalism 
is the one given by the optimized effective potential method
\cite{Tal76}. 
It is known, however, that it is nearly the same as that given by
the HF calculation for atoms\cite{Tal76,Li93}. 

We first compare the results of HF, LDAX and SA-SICX 
for four different Na clusters.
The order of the estimated exchange energies, which are negative, is 
LDAX $>$ SA-SICX $>$ HF irrespectively of the size of the Na cluster.
This is different from the order for atoms, where 
LDAX $>$ HF $>$ SA-SICX \cite{Har83a}.
The deviation of the result of LDAX from that of 
HF gets smaller with the size of the Na cluster, while 
that of SA-SICX is almost constant. 

We next compare the results of HF and three SIC methods.
The absolute value of the exchange energy 
calculated by using the spherical harmonic and Cartesian orbital densities 
becomes smaller than that estimated by the SA-SICX.
Including the negative sign, the order is C-SICX $>$ SH-SICX $>$ SA-SICX.
Consequently, the deviation of the results of SH-SICX and C-SICX
from that of the HF gets larger than that for SA-SICX.
These two calculations are even worse than LDAX
for large systems. Their deviation from the HF calculation gets 
larger with increasing size of the cluster. 

For atoms, the order of the exchange energies 
C-SICX $>$ SH-SICX $>$ SA-SICX is the same as that for 
Na clusters. 
However, the result of C-SICX is still below that of HF.
This means that both of SH-SICX and C-SICX 
have smaller deviation from the HF result than 
SA-SICX and provide better prescriptions for atoms. 

\begin{table}
  \begin{center}
    \begin{tabular}[H]{ccccc} 
      &Na$_{8}$ &Na$_{20}$ &Na$_{40}$ &Na$_{92}$\\
      \hline
      {HF }
      & $ -2.95 $ & $ -2.95 $ & $ -2.95 $ & $ -3.03 $ \\ 
      {LDAX }
      & $ -2.68 $ & $ -2.78 $ & $ -2.84 $ & $ -2.93 $ \\ 
      {SA-SICX }
      & $ -2.93 $ & $ -2.91 $ & $ -2.91 $ & $ -2.95 $ \\ 
      {SH-SICX }
      & $ -2.87 $ & $ -2.84 $ & $ -2.83 $ & $ -2.87 $ \\
      {(H only)}
      & $ -3.00 $ & $-2.96  $ & $ -2.96 $ & $ -3.00 $ \\ 
      {C-SICX}
      & $ -2.82 $ & $ -2.78 $ & $ -2.77 $ & $ -2.80 $ \\
      {(H only)}
      & $ -3.07 $ & $ -3.01 $ & $ -3.00 $ & $-3.02 $ \\ 
    \end{tabular}
    \caption{Exchange energy per electron in eV.
      SA-SICX, SH-SICX and C-SICX are the abbreviations
      of the SICX calculated with spherically averaged,
      spherical harmonic and Cartesian orbital densities. 
      H only means that 
      only the Hartree term has been calculated 
      by using either spherical 
      harmonic or Cartesian bases (see text).}
    \label{tab:EX}
  \end{center}
\end{table}

\begin{figure}
  \begin{center}
    \epsfig{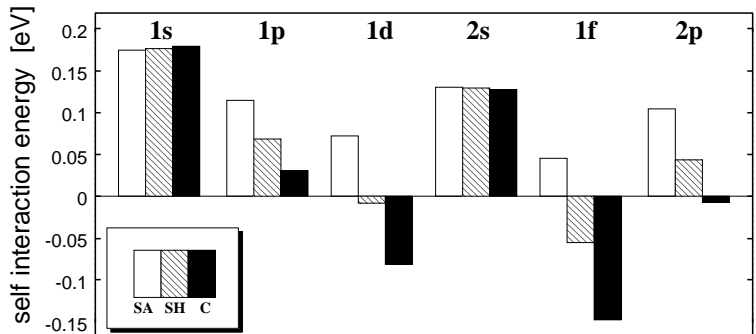}
    \caption{Each orbital contribution to the self interaction energy 
      for Na$_{40}$. SA, SH and C mean the SICX with spherically averaged, 
      spherical harmonic and Cartesian orbital densities, respectively.}
    \label{fig:orbital}
  \end{center}
\end{figure}

We now investigate the reason why the SH-SICX and C-SICX
methods are inferior to SA-SICX for Na clusters. 
Since the self-consistent density in the three SICX methods
is almost the same, the first term of Eq.(\ref{eq:mEX}) is 
nearly the same for the three methods. 
The major difference among them is therefore associated with 
the second term of Eq.(\ref{eq:mEX}), which consists of  
contributions from orbitals with various 
{$n,l$} quantum numbers. 
In order to see the physics clearly, 
we consider the following self interaction averaged over the 
azimuthal quantum number 
for each set of quantum numbers {$n,l$}, 
\begin{eqnarray}
  \frac{1}{2l+1}\sum_{m}\left\{
    E_{H}[\rho_{lnm}]+E_{X}^{\mathrm{LDA}}[\rho_{nlm},0]\right\} \ .
  \label{eq:SI}
\end{eqnarray}
Its value evaluated by inserting the orbital densities 
given by Eqs.(\ref{eq:rhoSA}), (\ref{eq:rhoSH}) and (\ref{eq:rhoC}) 
is compared in Fig.\ref{fig:orbital} for the occupied orbitals in 
Na$_{40}$.

The three methods agree well for the $s$ orbitals.
The small differences originate from a subtle difference 
of the total densities in the three methods.
On the other hand, a large difference appears among them 
for the finite angular momentum states, \textit{i.e.} 
for the $p$, $d$ and $f$ orbitals. 
Two specific features can be remarked, 
i) always SA-SICX $>$ SH-SICX $>$ C-SICX, and 
ii) the higher the orbital angular momentum is, the larger the differences
among them are.

The key point to understand these features is the degree of 
localization of single particle density adopted in the 
three SICX methods. 
Clearly they differ in their angular treatment 
of the orbital density. 
C-SICX has the most distinct localization in angle,
while SA-SICX has, of course, no angular localization. 
Symborically, we express this situation of the degree of localization 
as C-SICX $>$ SH-SICX $>$ SA-SICX.  
This difference gets more prominent for higher angular momentum.

The problem is that the angular localization property affects 
the self Hartree term, \textit{i.e.} the first 
term in the curly brackets in Eq.(\ref{eq:SICX}),
and the self exchange terms, \textit{i.e.} the second term, in different way. 
Since the self Hartree term uses the 
exact long range Coulomb interaction between electrons, 
it does not depend so much on the 
density profile of electrons including the 
angular localization. 
On the other hand, the self exchange term is evaluated based on the LDA.
This is equivalent to assuming a short range 
$\delta$ interaction, 
which leads  to a strong sensitivity 
of the exchange energy on the details of the density profile of 
electrons.
The more the orbital localizes, the larger the absolute value 
of the self exchange energy is. 
In C-SICX, which has the most prominent angular localization, 
the negative self exchange energy increases with angular momentum 
and eventually overwhelms the self Hartree term resulting in the sign 
change of the total self interaction energy. This can be clearly seen 
in Fig.\ref{fig:orbital}. A less prominent, but a similar, trend 
can be seen also for SH-SICX. 
However, this strong sensitivity 
of the self correction energy to the 
angular localization of electronic density
may be unphysical because the LDA for the exchange term 
cannot be justified for the localized density in SH-SICX and C-SICX. 
The use of spherically averaged density in the SA-SICX 
moderates to some extent the 
overamplification of the angular localization dependendence 
leading to a better approximation. 

The localization of radial wave functions in atoms is very different from
that in Na clusters when one compares the states with the same nodal quantum 
number. For example, $1s$ and $1p$ orbitals have very similar radial distibution 
in Na clusters, while the latter ($2p$ in the atomic notation) 
is much more extended than the former in atoms. 
Consequently, the self interaction correction mainly originates from
the most localized $1s$ state\cite{Per81} in atoms.
Moreover, only small angular momentum orbitals
appear in atoms. Thus, the problem stated above, \textit{i.e.} the unphysical 
strong angular localization dependence does not cause so serious 
trouble in atoms. 

To remedy the problem for Na clusters, we propose to 
utilize spherical harmonic and
Cartesian orbital densities only for the self Hartree term.
This is a natural consequence of the considerations mentioned above. 
The exchange energy calculated in this way 
is given in the lower side in the 
rows for SH-SICX and C-SICX in Table \ref{tab:EX}.
They are designated by the label ``(H only)''.
All of the \textit{H-only} exchange energies 
better agree with that in HF than 
the corresponding results of the SH-SICX and C-SICX which 
have been calculated using the spherical harmonic or
Cartesian orbital densities both for the self Hartree and exchange terms 
(the upper side in each row).
Especially, \textit{H-only} SH-SICX is superior to
the SA-SICX for most systems.
We remark that a similar improvement has been obtained by the 
\textit{H-only} method concerning the total energy.   

\begin{table}
  \begin{center}
    \hspace*{-2.5cm}
    \begin{tabular}[c]{ccccc} 
      & Na$_{8}$ & Na$_{20}$ & Na$_{40}$ & Na$_{92}$ \\ \hline 
      HF        &$ -140.9 $&$ -626.2 $&$ -1957.6 $&$ -7766.7 $ \\ 
      LDAX      &$ -138.9 (+1.42\%) $&$ -623.0 (+0.51\%) $&$
      -1953.3 (+0.22\%)$&$ -7758.1 (+0.11\%)$ \\ 
      SA-SICX   &$ -140.8 (+0.07\%)$&$ -625.7 (+0.08\%) $&$
      -1956.4 (+0.06\%)$&$ -7761.1 (+0.07\%) $\\ 
      SH-SICX   &$ -140.4 (+0.35\%)$&$ -624.4 (+0.29\%)$&$
      -1953.5 (+0.21\%)$&$ -7754.0 (+0.16\%)$ \\ 
      (H only)  &$ -141.3 (-0.28\%)$&$ -626.8 (-0.10\%)$&$
      -1958.4(-0.04\%) $&$ -7765.1(+0.02\%)$ \\ 
      C-SICX    &$ -140.1 (+0.57\%)$&$ -623.2 (+0.48\%)$&$
      -1951.0 (+0.34\%)$&$ -7747.7 (+0.24\%)$ \\ 
      (H only)  &$ -141.8 (-0.64\%)$&$ -627.7 (-0.24\%)$&$ 
      -1960.0 (-0.12\%) $&$ -7767.5 (-0.01\%) $\\ 
    \end{tabular}
    \caption{Total energy for Na clusters in eV.
      The relative error of each method
      (the difference between the values in each method and HF
      divided by the HF value) is shown in parentheses.}
    \label{tab:ETOT}
  \end{center}
\end{table}

\subsection{Effects of non-diagonal Lagrange multipliers}
\label{sec:ortho}
Before we conclude the paper we comment on the effects of 
off-diagonal Lagrange multipliers in Eq.(\ref{eq:1eq}). 
We performed calculations by keeping off-diagonal Lagrange multipliers
for Na$_{20}$ and Na$_{40}$ with SA-SICX, SH-SICX and C-SICX.
We found that the differences of the total energy and single particle 
energy obtained by self-consistently solving 
Eq.(\ref{eq:1eq}) and Eq.(\ref{eq:diag1eq})
are less than  10$^{-1}$ eV and  10$^{-2}$ eV, respectively.
We therefore conjecture that one can safely 
ignore the off-diagonal components of the Lagrange multipliers
for Na clusters.
This result is consistent with that in Ref. \cite{Har83b} for 
atoms. 

\section{Summary and conclusion}
\label{sec:summ}
We have calculated the self interaction corrected exchange energy
for Na clusters by using spherically averaged, spherical harmonic
and Cartesian orbital densities.
We found that both calculations using the spherical harmonic and
Cartesian orbital densities deviate from the HF results
more than the calculations with spherically averaged orbital densities.
The deviation is especially large for systems
with large angular momentum orbitals.
We attribute this problem to the LDA to evaluate the self exchange term. 
From this consideration, we propose to use the spherical harmonic and
Cartesian orbital deinsities only for the self Hartree term and to use 
spherically averaged orbital densities for the self exchange term.
We have shown that this treatment improves indeed 
the exchange energy to well reproduce the results of HF calculations.
We expect that the remaining errors can be diminished by the SIC
using GGA(generalized gradient approximation) 
\cite{Per86,Bec88,Per91,Ish95}. 

\section*{Acknowledgement}
We are thankful to David M. Brink for enlightening discussions.

\end{document}